\documentclass[a4paper]{jpconf}
\usepackage{graphicx}
\usepackage{color}

\newcommand{\et}{{\it et~al.}}
\newcommand{\eg}{{\it e.~g.}}

\begin{document}
\title{Solar extreme events}

\author{Hugh S. Hudson}
\address{Space Sciences Laboratory, University of California, 
7 Gauss Way, Berkeley CA 94720-7450, USA}

\ead{hhudson@ssl.berkeley.edu}

\begin{abstract}
Solar flares and CMEs have a broad range of magnitudes.
This review discusses the possibility of ``extreme events,'' defined as those with magnitudes greater than have
been seen in the existing historical record.
For most quantitative measures, this direct information does not extend more than a century and a half into
the recent past.
The magnitude distributions (occurrence frequencies) of solar events (flares/CMEs) typically decrease with the parameter measured or inferred (peak flux, mass, energy etc.
Flare radiation fluxes tend to follow a power law slightly flatter than $S^{-2}$, where S represents a peak flux; solar particle events (SPEs) follow a still flatter power law up to a limiting magnitude, and then appear to roll over to a steeper distribution, which may take an exponential form or follow a broken power law.
This inference comes from the terrestrial $^{14}$C record and from the depth dependence of various radioisotope proxies in the lunar regolith and in meteorites.
Recently major new observational results have impacted our use of the relatively limited historical record in new ways: the detection of actual events in the $^{14}$C tree-ring records, and the systematic observations of flares and ``superflares'' by the Kepler spacecraft.
I discuss how these new findings may affect our understanding of the distribution function expected for extreme solar events.
\end{abstract}

\section{Introduction}
Flares and coronal mass ejections pose well-recognized hazards for human society on Earth and its outposts in space.
The damage inflicted by extreme solar events,\footnote{This paper adopts the working definition of ``extreme event'' as one more powerful than  any heretofore observed; see Section~\ref{sec:what} for discussion.}
also known as ``superflares'' and perhaps ``super CMEs'' will by definition exceed the levels suffered in the historical past.
Mission planning for manned space flight, for example, requires adequate knowledge of the fluence distribution in Solar Particle Events (SPEs) and the likelihood of the occurrence of such events.
Our knowledge of the distribution function in magnitude of the events generally points to scale-free power laws  (e.g., \cite{1991SoPh..133..357H}), as derived from direct measurements of particle or photon fluxes.
In this paper we point out progress made in recent years that adds new proxy information that may augment the direct measurements.
This new information in principle could change the assumptions needed for forecasting the occurrence of an extreme event, and may eventually make predictions qualitatively more reliable.

The approximately power-law distribution of flare peak fluxes has been known since the Van Allen era of the space age \cite{1971SoPh...16..152D} or even before \cite{1956PASJ....8..173A}.
Note that these papers deal with soft X-ray and microwave bursts, respectively, but observations in many other spectral regions show the same property.
Thus we often assume that this power-law characteristic refers basically to the flare energy, as seen through various proxies (such as these), even though any one of them may deviate systematically from this fundamental property.
As is well known, such featureless power-law distributions appear in many other situations in nature -- Zipf's Law; the Gutenberg-Richter Law; Paneto's Law, etc. (\eg, \cite{2013socs.book.....A}).

The particle data come directly (at low cosmic-ray energies) from counters deployed in interplanetary space.
Van Hollebeke \et~\cite{1975SoPh...41..189V} presented an early comprehensive study of fluence distributions for energetic protons in the range 20-80~MeV, finding a power-law fit for maximum particle intensity $I$ (particles~cm$^{-2}$~s$^{-1}$~sr$^{-1}$~MeV$^{-1}$) of the form $dN/dI \sim I^{-\alpha}$, with $\alpha = 1.15 \pm 0.1$.
The modern data also include ground-based observations from neutron monitors (e.g., \cite{2000SSRv...93...11S}).
These respond at higher energies and so do not directly measure the most damaging components of the particle spectrum; the neutron monitors also only detect the most powerful events.
On the other hand, it is just these extreme events that  dominate the particle fluence, and so the information the neutron monitors provide is essential.

The flat power-law distribution of peak particle intensities, carried over to a distribution of fluences, poses a problem for predictions. 
Such a flat power law leads to uncertainty in estimates of fluences at the top end of the scale; for a flat distribution, the one or two greatest events in a given solar cycle may well provide the bulk of the particle fluence for that entire cycle.
The question of occurrence probability then requires an understanding of the likelihood of an event with as great a fluence as any observed, or a still greater one of a magnitude not yet directly recorded, and yet still permissible theoretically.

Modern data have greatly improved our knowledge of the non-extreme events, and one objective of this paper is to put this new knowledge in the context of predictions of occurrence probability.
Schrijver \et~\cite{2012JGRA..117.8103S} give the starting point for the present review; this paper surveys flare energetics, and notes the strong evidence for the existence of an upper cutoff in total event energy.
Flares (and CMEs) do not directly scale with solar particle events (SPEs); the relationships are strong but not simple.
In this review we start by discussing the basically new information from tree rings and from the Kepler stellar observations (Section~\ref{sec:new}).
Section~\ref{sec:direct} describes the bearing of this information on the relationship between SPE magnitude to total event energy, or flare magnitude.
This inspires a ``search for the break'' (Section~\ref{sec:break}), assessing our current knowledge of behavior of the occurrence distribution beyond our direct empirical knowledge.
Section~\ref{sec:extreme} attempts to reconcile these disparate observations, emphasizing the importance of the rollover to large event magnitudes.
The existence of this break energy would basically resolve the prediction issue posed by the flat distribution function apparent in the data of Van Hollebeke et al. \cite{1975SoPh...41..189V}, for example, in that integration of the distribution now can lead to a reduction in our best guess about the occurrence probability of an extreme solar event.

\section{What is an extreme event?}\label{sec:what}

The flat power-law distributions of observed flare magnitudes and particle fluences mean that the major events dominate the total within the distribution, which in statistical jargon has a ``heavy tail.''
These observations therefore provide an antithesis of the ``nanoflare'' hypothesis \cite{1988ApJ...330..474P}, in which the accumulation of small events dominates the total.
From other considerations an extreme solar event would be one that could do far more damage to our environment than ones previously experienced; such an event would have a vanishing probability of occurrence until it actually happened and thus instantly became an unavoidable reality.
Thus the manner in which we extrapolate our observational experience can have a profound influence on assets with vulnerabilities; on an arguably small scale one can point to power-grid failures and satellite outages as costly and unexpected failures of large-scale assets (e.g., \cite{2014GeoRL..41..287T}) on Earth and in the near-Earth space environment.

The main immediate interest in damge produced by solar extreme events is in the behavior of the solar energetic particles (SEPs or ``solar cosmic rays; see e.g. \cite{2008IJMPA..23....1M,2013SSRv..175...53R} for recent reviews).
We must bear in mind, however, that our paradigm derived from weaker events may not apply accurately to much stronger ones; in addition to the SEPs there might also be substantial impact from hard electromagnetic radiations (EUV, X-rays, $\gamma$-rays) or from destabilization of the Earth's magnetosphere by the impact of accompanying interplanetary CME.

The Carrington flare (SOL1859-09-01), at the very boundary of the historical record, could have approached  the actual occurrence of a solar extreme event \cite{2003JGRA..108.1268T}, in terms of later experience, although this interpretation as the source of an extreme geomagnetic storm may have been based on a misunderstanding of the geomagnetic signatures recorded at low latitudes \cite{2006AdSpR..38..173S,2013JSWSC...3A..31C}.
The flare and inferred CME did not greatly exceed, by most measures, other more recent and better-studied major events such as SOL2003-11-04 \cite{2013JSWSC...3A..31C}, illustrated below, or SOL2012-07-23 \cite{2013SpWea..11..585B}. 
So no clear evidence for a ``Black Swan''\footnote{A ``Black Swan'' is extreme event \cite{taleb2007the}, in popular terminology now adopted by some disaster planning offices. ``Such rare events are observed in exceptionally strong windgusts, monster or rogue waves, earthquakes, and financial crashes'' \cite{2014FoPh...44..546K} -- and possibly in stellar magnetic activity as well}.  
seems to exist on a time scale now exceeding one century.

\section{Newer developments}\label{sec:new}

\subsection{Events discovered via analysis of tree rings}\label{sec:rings}

Living things ingest carbon, and this contains a time stamp in terms of the immediate abundance of the $^{14}$C radioisotope.
Its radioactivity then lets us date a biological sample, a property used in many studies for different purposes.
Prior to the publication of the first direct event detection via the $^{14}$C proxy in tree rings \cite{2012Natur.486..240M}, we had only statistical information about the time dependence of the production rate of the $^{14}$C isotope.
This first event, which occurred in about 775~AD based on the tree-ring chronology (``dendrochronology''), has generated substantial discussion and certainly could have been solar in origin.
The same group soon uncovered a second event \cite{2013NatCo...4E1748M}, dated about 993~AD.
The detection of these events required the improvement of time resolution for the radioisotope signature to about one year, as compared with the few-year
resolution previously available.
The time resolution of such observations, in terms of $^{14}$C rate, suffers because of the complicated processes involved in circulation and uptake into the trees; this of course limits the degree to which such an event could be linked to some other phenomenon.
The initial discovery of the event was in annual samples of cedar trees from Yaku Island, Japan, and its presence was soon confirmed in annual data obtained from trees in Germany, Austria, the U.S., Russia, and New Zealand, as well as via $^{10}$Be records from Antarctica \cite{2013A&A...552L...3U,2014GeoRL..41.3004J} and Chinese corals \cite{2014NatSR...4E3728L}.
This widespread effect provides further evidence for an extraterrestrial origin.

\begin{figure}[htbp]
\noindent\includegraphics[width=19pc]{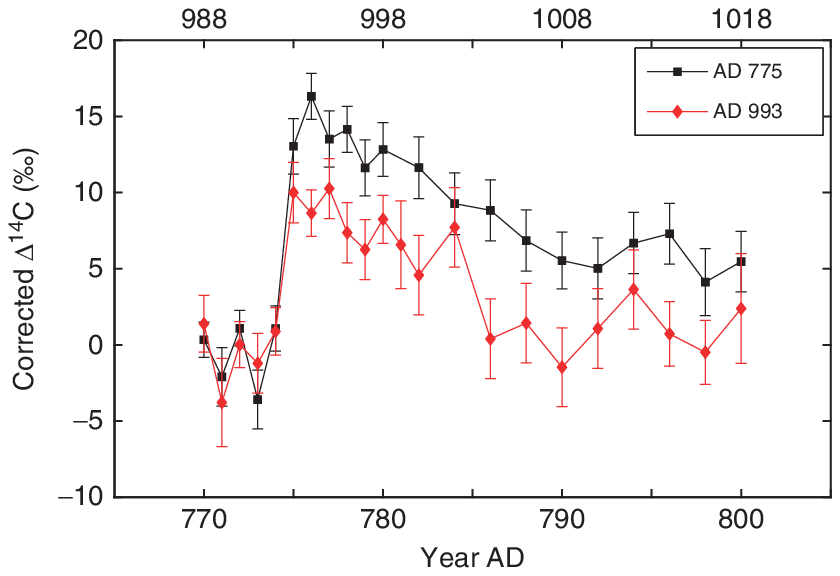}
\noindent\includegraphics[width=18pc]{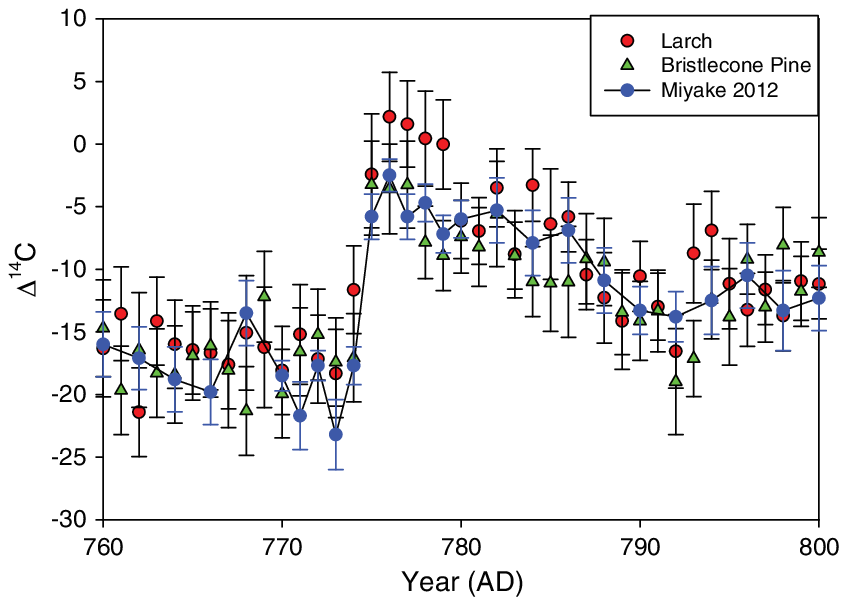}
 \caption{Left, the two tree-ring radioisotope events from Japanese cedars, from annual samples \cite{2013NatCo...4E1748M}.
 At this resolution these time series each show an unresolved sudden increase, followed by a slow recovery consistent with terrestrial transport and the uptake of 
 $^{14}$C into the tree.
 Right, earlier event also appeared in tree rings from larches, kauri, bristlecone pines, and oaks, and from both N and S hemispheres \cite{2014GeoRL..41.3004J}.
 }
 \label{fig:miyake_two_events}
 \end{figure}

The tree-ring records have provided our first radioisotope evidence for discrete high-energy transients in the terrestrial atmosphere.
Could the Sun have been responsible \cite{2013A&A...552L...3U}?
Similar increases cannot be detected positively in the isotope records for known major flare events, such as SOL1859-09-01 (the Carrington event), and
so this would imply that the two new events fall in a new hypothetical category of solar ``superflares.''
The significance of these events would then be that they constrain the distribution of solar events on longer time scales than the extent of the historical record, and thus help define the probability of occurrence of less frequent but more powerful events.
Even if a non-solar explanation prevails \cite{2014ApJ...781...32C,2014AN....335..949N,2014AstL...40..640P}, the absence of other detections still provides upper limits for solar event occurrence on the relevant time scales.
The tree-ring events could be consistent with a solar origin, given the roll-off and with an appropriate spectral distribution \cite{2012ApJ...757...92U,2014ApJ...781...32C}, but would reflect a much larger event than the F$_{30}$ cutoff discussed below in Section~\ref{sec:spes}; here F$_{30}$ refers to the fluence of solar particles above 30~MeV, a standard categorization.
The cutoff occurs near F$_{30} = 10^{10}$~protons~cm$^{-2}$, but a proton energy of 200~MeV may be more appropriate to allow for reduced model-dependence on the spectral distribution, especially for the $^{10}$Be isotope \cite{2014SoPh..289.4691K}.
The rare appearance of such a solar event, over such a long time span, might not conflict with our interpretation in terms of the steepening of the distribution Section~\ref{sec:break}.

\subsection{Stellar flares}\label{sec:stellar}

Stars of many spectral classes, but most obviously the UV Ceti ``flare star'' category (dMe), produce flares that more or less match the paradigms of solar flares (e.g., \cite{2005stam.book.....G}).
In general we lack the breadth of diagnostics for these stellar flares that we have for the solar ones, and in particular we do not have systematic observations of hard X-rays and $\gamma$-rays that could reveal analogous particle acceleration -- this is a matter of sensitivity.

\begin{figure}[htbp]
\noindent\includegraphics[width=19pc]{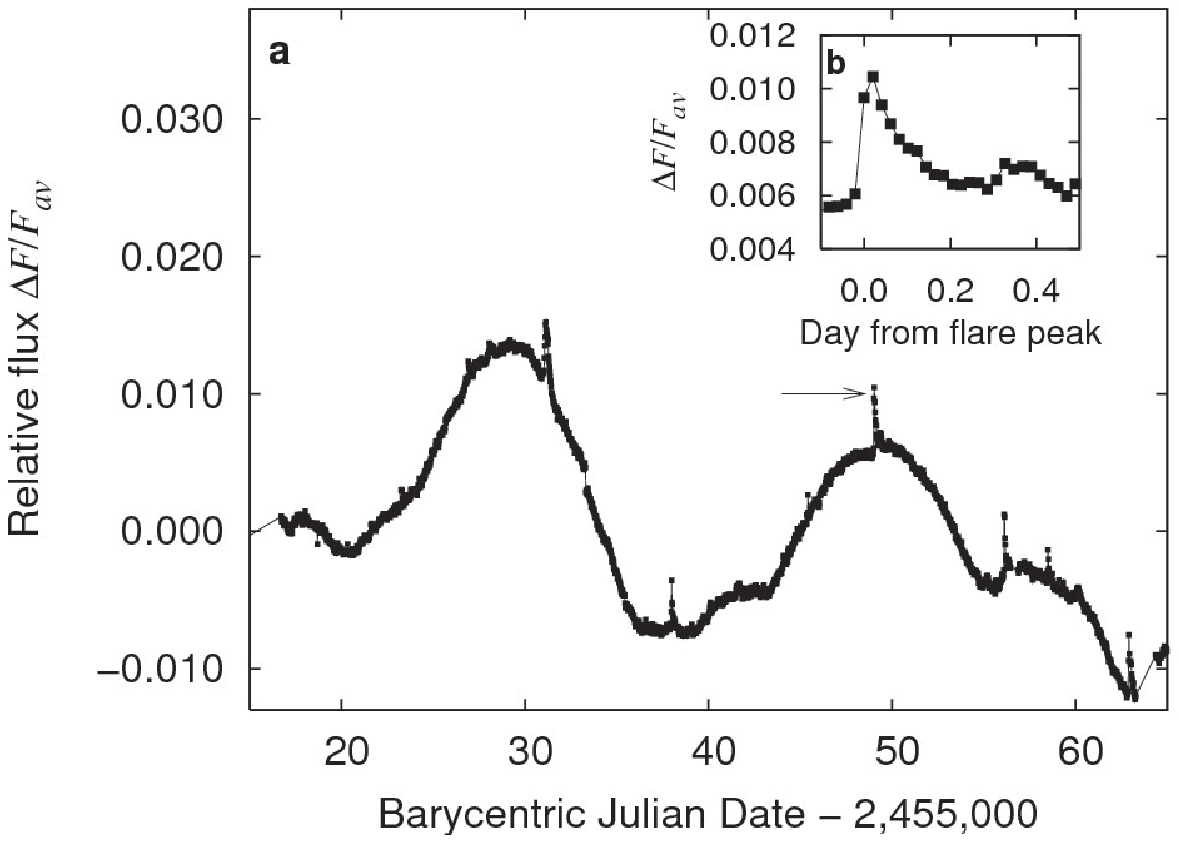}
\noindent\includegraphics[width=19pc]{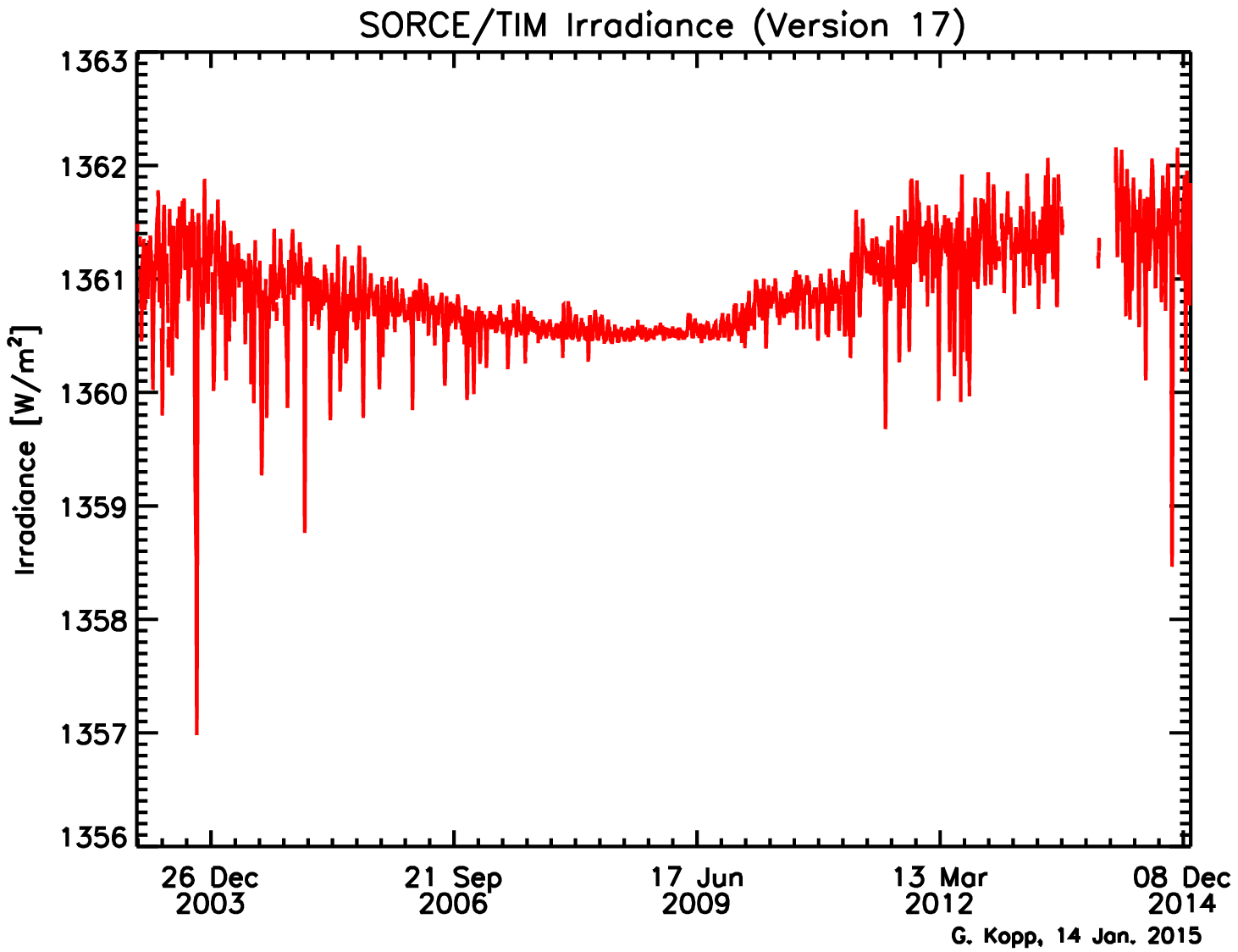}
 \caption{Left, flaring observed by \textit{Kepler} on the slowly-rotating G-type main sequence star KIC~11764567 \cite{2012Natur.485..478M}; (a) shows a 50-day time series with slow modulations attributed to large starspots 
 (e.g., \cite{2005LRSP....2....8B}), and the inset (b) for a fraction of a day.
Right, a full 11-year cycle of solar bolometric data (courtesy G. Kopp; for an early view of such solar variability via the \textit{Solar Maximum Mission's} ACRIM radiometer, see \cite{1981Sci...211..700W}).
 Here the background variability has much shorter time scales and much smaller amplitude; there are no flare-related increases \cite{1983SoPh...86..123H} and instead, pronounced ``dips'' from individual large sunspot groups.
 Note the contrast between these rapid dips (about 1/4 of the rotation period, and not sinusoidal) and the slower and larger modulations of KIC~11764567. 
 The photometric behavior of these two stars could hardly be more different.
 }
 \label{fig:maehara_kepler}
 \end{figure}

Recently the \textit{Kepler} satellite has opened a database on precise stellar time-series photometry with enormous qualitative and quantitative improvements  \cite{2010ApJ...713L..79K} in the statistics of solar-type flare occurrence on other stars.
The new data reveal myriads of flares \cite{2011AJ....141...50W,2012MNRAS.423.3420B}, including ``superflares'' on solar-type stars \cite{2012Natur.485..478M,2013ApJS..209....5S,2015ApJ...798...92W} that may help us to understand extreme solar events.
Figure~\ref{fig:maehara_kepler} shows an example of one of the flaring solar-type \textit{Kepler} stars.
The large, slow variations (of order 1\%) suggest the presence of large-scale magnetic structures on the stellar photosphere (solar-type spots?); superposed on these variations one sees the impulsive increases of the stellar flares.

How similar are the variations seen in the \textit{Kepler} stars and the Sun itself?
We note the contrast in the behavior of the two stars compared in Figure~\ref{fig:maehara_kepler} (KIC~11764567, via broad-band photometry, and the Sun as viewed in total solar irradiance).
The  \textit{Kepler} observations directly suggest rotational modulation with a simple structure, whereas the solar timeseries has more complexity.
The solar spots appear to be far more concentrated, and appear in the timeseries as downward excursions lasting for about a quarter of the rotational period in a distinctly non-sinusoidal manner (the ``dips'' \cite{1981Sci...211..700W}).
The distinction between these two morphological patterns probably does not depend upon the inclination of the stellar rotation axis, since exact modeling based on known sunspots shows little effect in the solar case for out-of-the-ecliptic viewing \cite{2012GeoRL..3916104V}.
From the Figure, it is difficult to understand how the solar paradigm applies to the stellar case, given the distinctly different morphologies.

\begin{figure}[htbp]
\noindent\includegraphics[width=30pc]{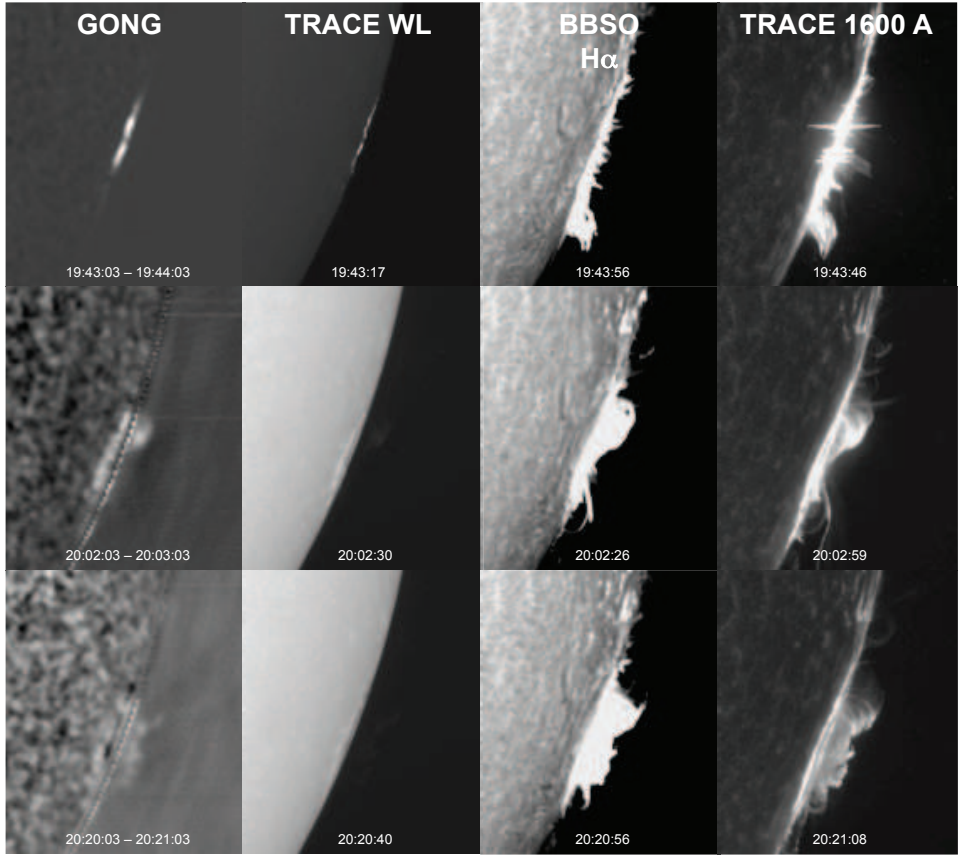}
 \caption{Observations of the limb flare SOL2003-11-04, which saturated GOES at the X17 level.
 From left to right, the columns show GONG narrow-band continuum;  TRACE white light, H$\alpha$, and TRACE 1600~\AA.
 The GONG signal has a background profile subtracted, and the above-the-limb portion amplified 10$\times$ (adapted from \cite{2004AAS...204.0213L}).
 }
 \label{fig:wlp}
 \end{figure}

How well do the  \textit{Kepler} superflares describe what the Sun might do, in the case of an extreme event?
The Kepler photometry has only one broad passband (approximately 400-900~nm) and poor time resolution (typically 30~min, though sometimes 1~min), but excellent photometric precision \cite{2011ApJS..197....6G} and therefore sensitivity.
The broad passband for this photometry incorporates many chromospheric emission lines, in particular H$\alpha$, but also could include continuum analogous to that of a solar white-light flare.
The UV extension of this continuum, in the case of a solar flare, may not make an order-of-magnitude difference in estimates of a flare's radiant energy, but this conclusion is based mainly on the observations of total solar irradiance (TSI) in a very few flares \cite{2004GeoRL..3110802W,2006JGRA..11110S14W}.
If we make the assumption that the continuum (``white light flare'') component dominates both solar and Kepler flare signals, then we meet a morphological
discrepancy: solar white-light flares, including those detected via TSI \cite{2004GeoRL..3110802W,2006JGRA..11110S14W,2011A&A...530A..84K}, do not have such long durations.
On the Sun, true white-light flares, as seen in visible continuum emission, match the impulsive-phase time profile of hard X-rays well, usually not more than ten minutes in duration.
In rare cases (SOL2003-11-04), as seen in Figure~\ref{fig:wlp}) the emission can continue for tens of minutes, but a solar white-light flare with the duration of the flare seen in the  inset of Figure~\ref{fig:maehara_kepler} has never been reported.
We note the possibility that the Kepler flares shine by scattered light, a rare phenomenon in the solar case sometimes referred to as a ``white light prominence'' \cite{1983SoPh...86..185H,2004AAS...204.0213L,2014ApJ...780L..28M,2014ApJ...786L..19S}.
This radiation presumably results from Thomson scattering in coronal material ejected by the flare process from the lower atmosphere, and trapped in large-scale loop systems that extend visibly above the limb of the star.
This continuum, perhaps the source of the extended radiation from SOL2003-11-04, persisted for tens of minutes.
It would be easy to imagine a larger-scale version of this in a stellar flare, in which an ejected mass of order 10$^{18}$~g might produce a Thomson-scattered
photospheric signature competitive with the direct continuum emission.
Via this mechanism, the total energy required for a Kepler superflare might turn out to be less than previously estimated \cite{2012Natur.485..478M,2013ApJS..209....5S}.

 If we disregard these reservations and simply accept the Kepler superflares as an extension of the solar statistics, we find a distribution function roughly similar to that of solar flares, with a power-law index near 2 \cite{2012Natur.485..478M,2015ApJ...798...92W}, and an occurrence rate of one event per 800-5000 years at a total energy of 10$^{34-35}$~erg \cite{2013ApJS..209....5S}.
 Adopting a solar rate of roughly one Carrington event per 80-year interval \cite{2012SpWea..1002012R}, this seems about right for an integral probability scaling roughly as 1/energy (equivalent to $\alpha = 2$ in the differential power law; see Section~\ref{sec:flares}).
 This would conflict with the solar requirement for a roll-off to greater magnitudes (Section~\ref{sec:break}), but only weakly (Section~\ref{sec:extreme}).

Finally, do we have any theoretical guidance about whether the Sun can make superflares?
The answer to this question is a clear ``no'', except in the most general sense.
On the Sun, there is at most a weak correlation between spot area and flare magnitude (e.g., \cite{2000ApJ...540..583S}), so weak that it difficult to recognize even on solar-cycle time scales (e.g., \cite{2007ApJ...663L..45H}).
There is also virtually no correlation between inferred spottedness and Kepler flare magnitude (see Fig.~9 of \cite{2013ApJ...771..127N}, where the opposite conclusion is drawn from the observed scatter).
The magnetic morphology of the sunspots also plays a major, though ill-understood role \cite{1960AN....285..271K,2000ApJ...540..583S}.
In spite of this the existing theoretical ideas tend to emphasize the role of sunspot size \cite{1975ApJ...200..641M,2013PASJ...65...49S,2013A&A...549A..66A,2014ApJ...792...67C}, and the relatively smooth lightcurves of the Kepler superflare stars do suggest the presence of large-scale magnetic surface structures.
Nevertheless major H$\alpha$ solar flares can and do occur in spotless or minimally spotted regions \cite{1970SoPh...13..401D,1993SoPh..145..339S}, 
as can powerful X-class white-light flares (e.g., SOL1991-12-03 \cite{1992PASJ...44L..77H}).
Accordingly we should not attribute too much specificity to the relationship between the rotational modulation and superflare occurrence.

Nevertheless we might anticipate simple theoretical limits on the magnitudes of extreme events.
For example, the supergranulation scale \cite{2010LRSP....7....2R} has well-defined properties in the solar photosphere, and sunspot umbrae typically do not exceed this scale in area.
The coronal magnetic energy of this scale might be estimated from $W = \varepsilon L^2 H B^2/8\pi$,
where $\varepsilon$ would be the efficiency of field annihilation (say, 10\%), L the network horizontal scale (say, $3 \times 10^9$~cm), H the magnetic scale height (say, L/2), and B an estimate of the magnetic intensity of a maximal sunspot, say 5000~G.
Other rough estimates could be made, but this generous one gives $1.3 \times 10^{33}$~erg.
This would correspond to an event only an order of magnitude greater than that estimated for major flares 
or CMEs \cite{2012ApJ...759...71E}.
Another estimate could come from the total energy accumulated during a solar cycle's worth of dynamo action,
assuming that the coronal energy storage resets to zero at each solar minimum \cite{1979ICRC....5..323M}.
This argument suggests an upper cutoff at the observed magnitude of a once-per-cycle event, and
therefore not an extreme event.
If this reset of energy accumulation did not cease after each solar cycle, i.e. if the magnetic energy could be stored in the solar interior rather than in the corona, then the maximum energy of an event might increase proportionally with the accumulation time \cite{2013PASJ...65...49S} as needed to obtain an extreme event.

\section{Solar flares and solar energetic particles}\label{sec:direct}

We now turn to more direct information about event distributions, namely the solar record itself.
Solar extreme events can influence geospace via enhanced electromagnetic radiation, plasma flows, or high-energy particles.
These phenomena all have close relationships to solar flares (low corona) and/or CMEs (middle to high corona), and as a rule flares and CMEs occur in synchrony so close that distinguishing them physically, rather than observationally (flares occur in the solar atmosphere, CMEs in the corona), becomes difficult (e.g. \cite{2008ApJ...673L..95T}).
In near-Earth space (including the Moon), the solar energetic particles (SEPs, also known as ``solar cosmic rays'' \cite{2008IJMPA..23....1M}) provide the best proxy records and hence give us our best shot at characterizing the solar event distribution prior to the modern era.
Flares and CMEs have several clearly distinguishable kinds of particle acceleration: in the flare itself we have electrons at tens of keV detected via hard X-radiation in the flare impulsive phase, MeV-range electrons in flares associated with $\gamma$-ray emission \cite{2009ApJ...698L.152S}, the energetic ions of the flare $\gamma$-ray emission itself \cite{1995ApJ...455L.193R}, the long-duration high-energy $\gamma$-ray sources (\eg, \cite{1995ARA&A..33..239H} now observed by \textit{Fermi} \cite{2014ApJ...789...20A}; and then the particles underlying many coronal radio emissions \cite{1965sra..book.....K} and of course the SPEs themselves.
The distinct morphological properties of these high-energy signatures point to a variety of mechanisms for high-energy particle acceleration fundamental to both flares and CMEs.

We begin here with the distributions of flare radiation, and then contrast with the distributions of SEP events for which there is a proxy record.
One basic question relates to the nature of the relationship between the distributions of events of these very disparate types -- electromagnetic radiation from the base of the corona (or the photosphere itself), and particle fluxes closely related to the interplanetary medium at some great distance from the flare site, and with an ill-understood relationship to it.

\subsection{Distributions of flares}\label{sec:flares}

For any observable of a flare, and they extend from long radio wavelengths well into the $\gamma$-ray domain, one can construct a histogram of event numbers as a function of peak flux or time-integrated fluence.
For most of the 20th century, the H$\alpha$ observations defined flaring, but such ground-based observations presented obvious problems in terms of homogeneity. When routine radio observations became available, especially in the microwave range, it became clear that a simple power law could describe the occurrence distribution rather well \cite{1956PASJ....8..173A}.
Then we obtained access to space and could make X-ray observations, and these have now supplanted H$\alpha$ as the basic metric for flare occurrence ({e.g., \cite{2011SSRv..159...19F}).
The earliest long sequences of solar soft X-ray observations came from the Geiger-counter experiments of the Van Allen group, from which again power-law distributions provided good fits \cite{1971SoPh...16..152D}.
For the past few decades we have had a series of standardized X-ray photometers on board the GOES spacecraft (see \cite{2005SoPh..227..231W} for a recent description of these data); these systems have provided almost complete coverage with reasonably uniform calibrations for several decades (note that systematic microwave solar monitoring began in 1947 (e.g., \cite{1958JRASC..52..161C}), but with lower sensitivity relative to the soft X-ray signatures.
The flux levels from the GOES photometers unfortunately do not characterize the actual flare energy so directly; with the advent of TSI observations (as mentioned above) it turned out that the GOES soft X-ray energy amounts to less than 1\% of the total radiated energy \cite{2011A&A...530A..84K}.
The use of such a minor constituent as a measure of flare energy obviously could lead to confusion, and this applies pretty much to any individual observable except of course for the TSI, which unfortunately has extremely limited dynamic range -- only the most powerful events have been detected individually in this way.
Thus the theoretical desire to use the data to characterize total event energies basically has systematic problems, which however are becoming more tractable with modern data (e.g., \cite{2005JGRA..11011103E,2012ApJ...759...71E}).

The differential number $N$ of events at observed peak flux $S$,  per unit peak flux, follows $dN/dS \propto S^{-\alpha}$.
The slope $\alpha$ usually falls below 2.0; the original radio result from \cite{1956PASJ....8..173A} found a good power-law fit at 1.8, for example, and decades later a large sample of hard X-ray data led to a value $\alpha =  1.732 \pm 0.008$ \cite{1993SoPh..143..275C}.
Soft X-ray microflares also have flat distributions with $\alpha \approx 1.4-1.5$ as observed in \textit{Yohkoh} soft X-ray images \cite{1995PASJ...47..251S}.
The flatness of these distributions clearly points to a requirement for a break downwards at high peak fluxes, in order to avoid an infinity \cite{1991SoPh..133..357H}, as had been noted earlier for stellar flares \cite{1988A&A...205..197C}.
The flatness also implies that major events contribute as strongly as minor events (microflares or nanoflares \cite{1988ApJ...330..474P}) to the total energy budget of the process.

\subsection{Distributions of SPEs}\label{sec:spes}

The direct measurements tend to show power-law distributions of particle intensities or event fluences, as noted above.
Belov et al. \cite{2007SoPh..246..457B} find dN/dI~$\propto I^{-1.22\pm0.05}$ for well-connected GOES particle events $>$10~MeV (protons), and dN/dI~$\propto I^{-1.34\pm0.02}$ for all events.
Here $I$ denotes the peak particle intensity, often given in ``proton flux units'' (PFUs) of protons~(cm$^2$ s)$^{-1}$ above 1~MeV.
This confirms the original result of a very flat distribution by  \cite{1975SoPh...41..189V}, significantly flatter than even the flare distribution and therefore deeper still into the problem of convergence.
The flare radiative fluxes and SEP peak fluxes have very different distribution functions, which implies something about the physical mechanisms involved, since flares and SPEs can have such a close relationship.
Given this discrepancy, can we use the proxy records for SEPs to gain any insight into actual flare energies?

The observational situation has recently gotten some clarification: those flares associated with SPEs turn out to have a distribution function consistent with that of the SPEs themselves \cite{2012ApJ...756L..29C}, as shown in Figure~\ref{fig:cliver_dnds}.
The inference from this work is that more powerful flares have a greater propensity for particle acceleration, consistent with the finding that CME associations increase with flare magnitude (e.g. \cite{2006ApJ...650L.143Y}) and with the flatter distributions of the hard X-ray peak fluxes.
This distinguishes the SPEs qualitatively even at the flare site, establishing the requirement for eruption and hinting at a threshold effect for particle acceleration \cite{1978SoPh...57..237H}.
Note that this finding does not reveal the physics underlying this behavior, nor does it solve the problem of the divergence for extreme events.
We thus still have the problem of explaining a (dimensional) break point in the event distribution function.

\begin{figure}[htbp]
\noindent\includegraphics[width=18pc]{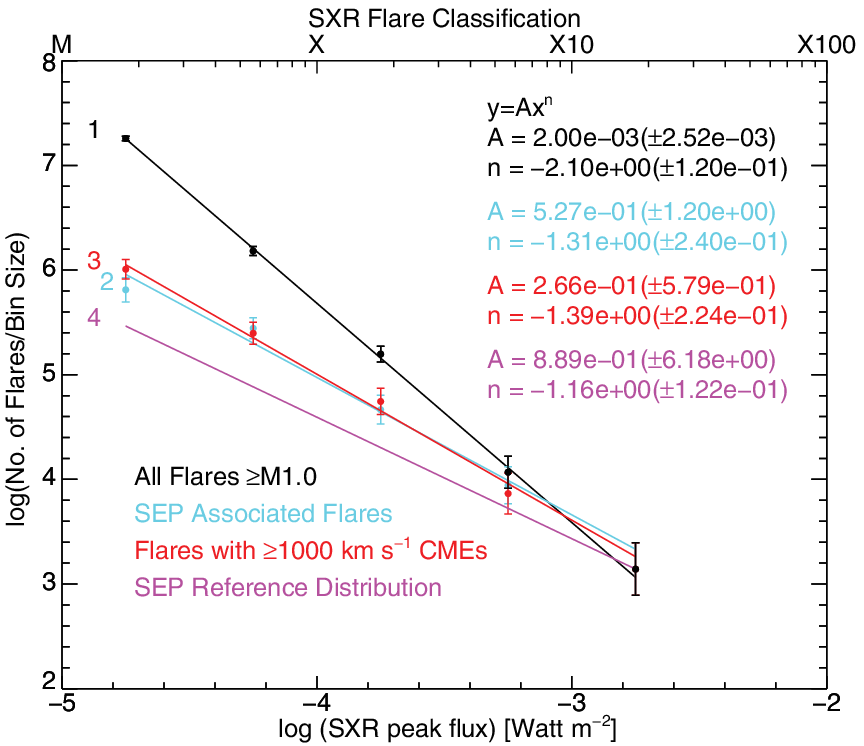}
\noindent\includegraphics[width=18pc]{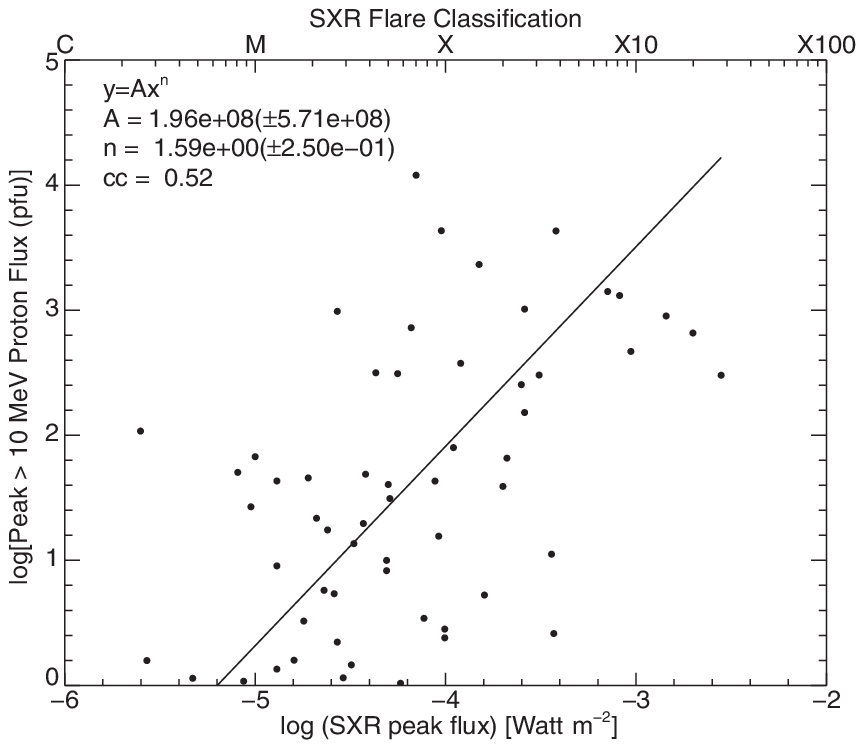}
 \caption{Left, distribution functions for GOES soft X-ray observations in the 1-8~\AA~band  (black) and SEPs (purple), showing the substantial difference between them. 
 Red and blue show flare distributions for the subset of events associated with SEPs and with fast CMEs.
 Note that the fit function in the figure legend defines our slope $\alpha$ as $-n$.
 Right, the correlation between peak proton and soft X-ray fluxes for the associated events  (figures adapted from \cite{2012ApJ...756L..29C}).
 }
 \label{fig:cliver_dnds}
 \end{figure}
 
\section{The search for a break}\label{sec:break}

No published dataset has yielded convincing evidence for a roll-over in Sun-as-a-star observations, for various reasons, but distributions constructed for flare occurrence in specific active regions via images may show such a phenomenon \cite{1997ApJ...475..338K}.

Even prior to the identification of discrete events in the tree-ring record (Section~\ref{sec:rings}), the fossil radioisotope record pointed clearly to the existence of a downward break in SEP event fluences \cite{1980asfr.symp...69L}.
Figure~\ref{fig:reedy} (left panel) shows an improved version of this result, from \cite{1996ASPC...95..429R}.
The isotope record includes both $^{14}$C from tree rings, and several other isotopes from the lunar regolith:
$^{41}$Ca, $^{81}$Kr, $^{36}$Cl, $^{26}$Al, $^{41}$Ca, $^{10}$Be, and $^{53}$Mn.
The half-lives of these isotopes range from $5.73 \times 10^3$ to $3.74 \times 10^6$ years, and these time scales individually establish effective integration times for each proxy isotope.

\begin{figure}
\noindent\includegraphics[width=20pc]{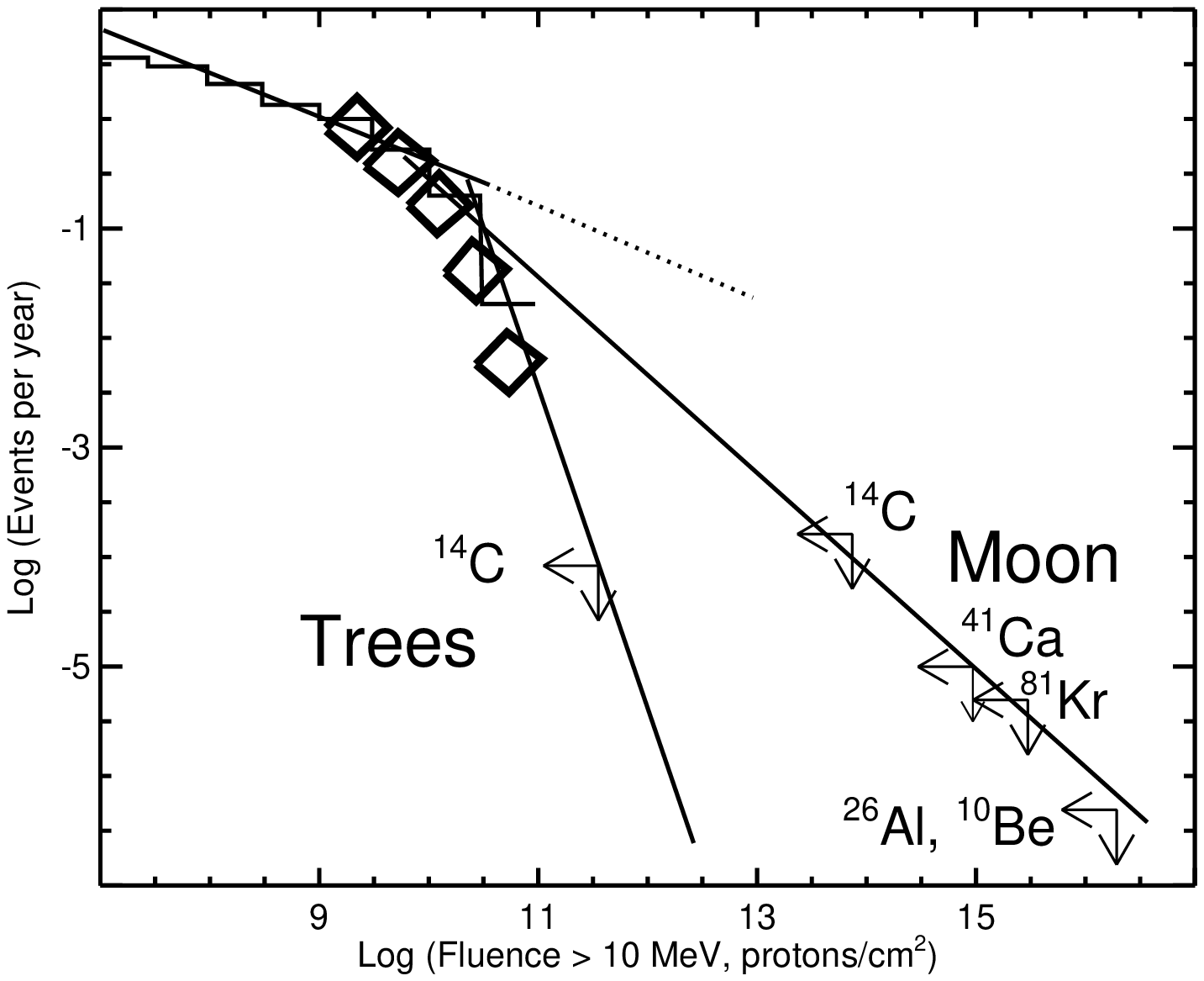}
\noindent\includegraphics[width=17pc]{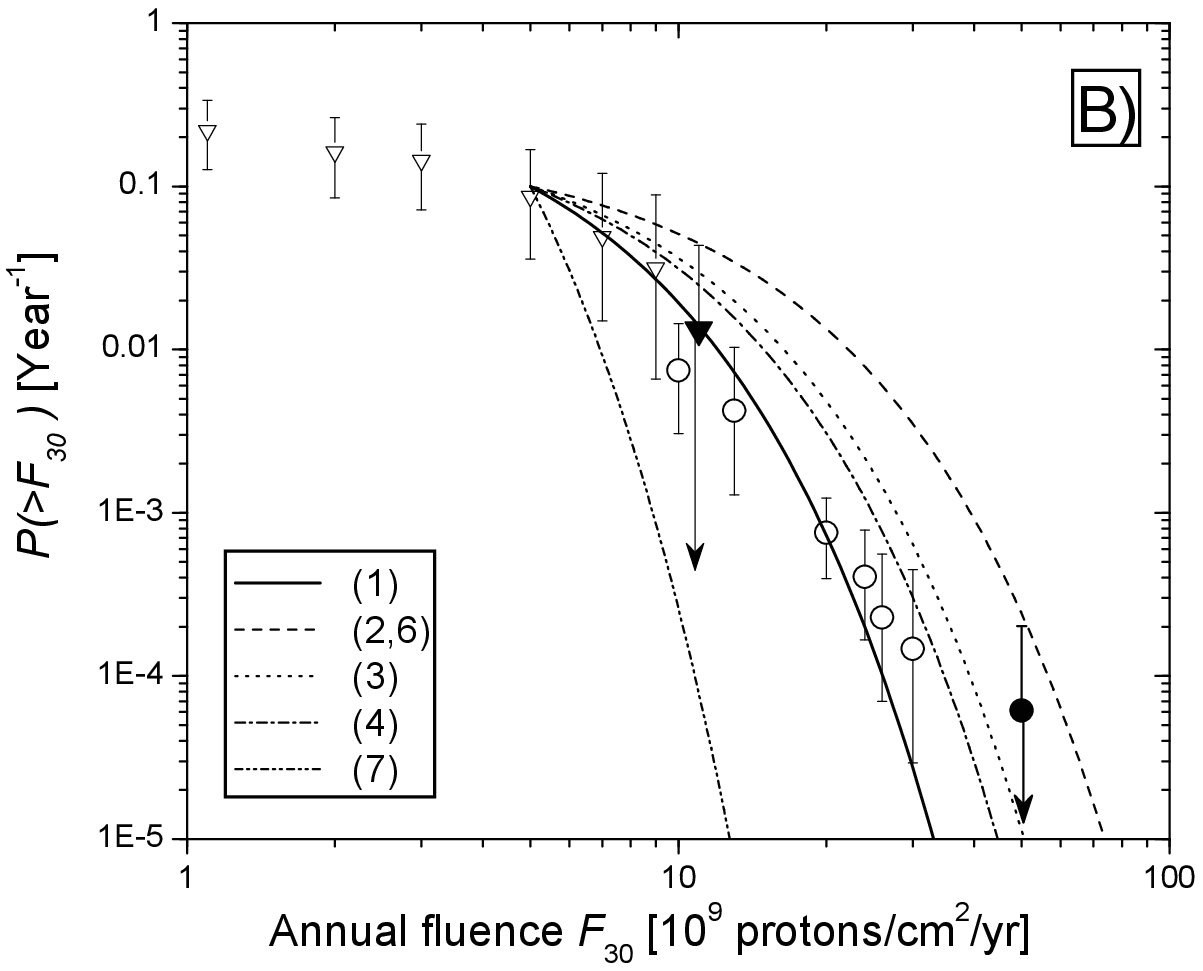}
 \caption{Left, the inference of the SEP fluence distribution from proxy radioisotope records (adapted from \cite{1996ASPC...95..429R}).
Right, an updated version of this information showing the terrestrial data as triangles and the limits for different isotopes according to their halflives as other symbols; the apparent break is at one event per decade at $5 \times 10^9$ protons (cm$^2$~yr)$^{-1}$ above 30~MeV (adapted from \cite{2014SoPh..289..211K}; see this paper for interesting details).
 }
 \label{fig:reedy}
\end{figure}

The right panel of Figure~\ref{fig:reedy} incorporates information from different isotopes from the lunar regolith, in the form of power-law functions consistent with a given isotope and anchored at a fixed reference point \cite{2014SoPh..289..211K}.
We identify the isotopes and the analyses as (1) $^{14}$C \cite{1998GeCoA..62.3025J}; (2) $^{41}$Ca \cite{1998GeCoA..62.2389F} and $^{26}$Al \cite{2001E&PSL.187..163G}; (3) $^{81}$Kr \cite{1999LPI....30.1643R}; (4) $^{36}$Cl \cite{2009GeCoA..73.2163N}; and (7) $^{10}$Be and $^{26}$Al from \cite{1996NIMPB.113..434M}.
Each of these papers described analyses of the depth dependence in lunar and meteoritic samples of the specific activities, often making use of accelerator experiments and numerical modeling to understand the cross-sections and the activation histories of the samples.

The downward break apparently occurs at a fluence of a few $\times 10^{10}$~protons/cm$^2$ ($>$30~MeV) \cite{2014SoPh..289.4691K}.
According to the results discussed above, this fluence could correlate with a specific GOES magnitude and total flare energy, but the physics of SEP fluence and flare radiation are so different that this poses problems theoretically as well as observationally.
As discussed above, the morphology rules out any identification of the SEPs and high-energy particles actually at the flare site, although we have not identified the acceleration mechanisms for the flare particles yet.
See \cite{2013SSRv..175...53R} for a recent systematic overview of the observational material on SEPs and the physical problems of their interpretation.
We observe the particles at a given point in space, or sometimes multiple points, but the particles that arrive at that point may have had a complex history of acceleration and propagation within the heliosphere, giving the appearance of diffusive storage in a heliospheric of a ``reservoir'' on a scale greater than one AU \cite{2013SSRv..175...53R}.
The existence of the reservoir may be inferred from the observations, but its understanding in terms of heliospheric field structure, even at the orbital distance of Jupiter \cite{2008ICRC....1..131S} remain unclear in detail.
Thus even the most direct measurements require model assumptions in their interpretation.
In terms of fluence, the physics of the diffusive shock acceleration theory (e.g., \cite{2013SSRv..175...53R}) also may play a confusing role, since the process may saturate and thus produce a nonlinear relationship (even if one were expected) in terms of flare energy, for example.
These uncertainties may contribute to the large scatter seen in the correlation of particle fluxes with flare magnitudes (Figure~\ref{fig:cliver_dnds}, right panel),
but the figure is consistent with a simple correlation.
In an interesting wrinkle here, the SEPs themselves (MeV particles) may contain an energy greater than that in the GOES soft X-ray sources, amounting in some cases to more than 10\% of the total flare energy \cite{2007AIPC..932..277M} rather than a small ``test particle'' fraction.

That the SEPs produced in a solar eruptive event may constitute a large fraction of the total event energy,
as estimated via flare radiation or CME kinetic energy, poses an interesting theoretical problem.
It encourages us to think of the event energy release as a proxy for an event's capacity for particle acceleration, but almost all existing theoretical work on flares and CMEs relies upon the single-fluid ideal MHD development.
In general there seems no alternative to the identification of the energy source with the coronal magnetic field, but its partition into the observable forms is known only from modeling \cite{2009ApJ...695.1151B} or, even more crudely, from the observations \cite{2012ApJ...759...71E}.
The information regarding any individual event is invariably incomplete, and the interpretation of any observable in terms of total energy remains entangled with uncertain and model-dependent assumptions.
Accordingly much work on occurrence distributions really deals with simple observables, such as the GOES soft X-ray flux or the peak particle intensity detected by a particular instrument, and an attempted conversion from any such measure to a theoretical construct often introduces new systematic errors.

\section{The occurrence of extreme events}\label{sec:extreme}

Can we do better than a simple extrapolation of the current data and obtain a better prediction of the occurrence of an extreme event?
See \cite{2012SpWea..1002012R} for a recent example of such a straighforward approach.
The proxy evidence on SEP fluences strongly suggests a downturn, and so a simple extrapolation of a flat power-law will definitely overestimate the probability.
The problem is that we have no precise information about the location of the break point in the distribution, nor any tangible hint that it might be detectable in the distribution of flare magnitudes (e.g., \cite{2009SoPh..258..141T}).
This Riley prediction   \cite{2012SpWea..1002012R} does not invoke the proxy information we have discussed in Section~\ref{sec:break}, instead relying simply on the extrapolation of the power law observable in the direct measurements into the realm of the extreme events.
This leads to a ten-year probability estimate of one 0.12~Carrington events per decade.
Including the rollover would reduce this probability by some factor that we has not been well quantified at present; note that either exponential or Band functions work equally well with the existing data \cite{2014SoPh..289..211K}.
This paper cannot really improve, therefore, on the 12\% solution for probability \cite{2012SpWea..1002012R}, except qualitatively.
The incorporation of a rollover in the distribution could reduce planning concern about a true Black Swan event, i.e. one of substantially greater magnitude than the Carrington event.

The new and interesting discoveries noted in Sections~\ref{sec:rings} and~\ref{sec:stellar} are as yet unable to help use substantially in defining the distribution of solar extreme events.
The two rapid $^{14}$C increases found in tree rings by \cite{2012Natur.486..240M,2013NatCo...4E1748M} have no corroborating evidence that would allow us to lock down the solar distribution except in an upper-limit sense; Miyake et al. and other authors had invoked various possible sources for these events, and we don't know for sure that they were solar in origin \cite{2012ApJ...757...92U}; on the other hand, from the point of view of astronauts or space hardware at risk, the origin of the events may not matter to first order.
These two events could in principle be evidence for a Black Swan phenomenon, describable by a power law without a termination, at some lower amplitude -- either solar, following a different paradigm, or non-solar in origin.

Similarly, the Kepler superflares may actually reflect a different paradigm, as seems to have been the case of the celebrated Elatina varves \cite{1985Natur.318..523W,1990RSPTA.330..445W} with their possible prehistorical record of the solar cycle.
In such a case, and as noted above there are striking dissimilarities between the Sun and the Kepler superflare stars, and we cannot blithely use them to assist in pinning down the actual solar event distribution (e.g., as described in \cite{2013PASJ...65...49S}), in spite of the ready availability of a solar analog \cite{2000ApJ...529.1026S,2011AJ....141...50W,2012Natur.485..478M}.
The finding of a correlation between starspots and flares in the Kepler data \cite{2013arXiv1304.7361N} would be consistent with standard ideas about magnetic energy storage and release in the corona \cite{2013A&A...549A..66A}, but the term ``superflare'' may connote the wrong paradigm.

A recent comprehensive discussion of the probability of an extreme event, identified with the Carrington flare and geomagnetic storm, Riley \cite{2012SpWea..1002012R},  concluded that 
``...our results overall suggest that the likelihood of another Carrington event occurring within the next decade is 12\%''
(for reference an independent assessment of essentially the same information led to an estimate corresponding to 3\% \cite{2012JGRA..117.8103S}, giving a feeling for the uncertainties here).
These analyses did not factor in the evidence for a rolloff to greater magnitudes above a certain limiting case, but this information can be incorporated into
a Bayesian estimation, for example. 
For example, following a method consistent with Bayesian or frequentist methods  \cite{2012GeoRL..3910301L} these two events would obtain 0.016 events per decade, with a 2$\sigma$ uncertainty range of [0.003, 0.09] events per decade.

\section{Conclusions}
 The occurrence of solar extreme events, by definition those flares and/or CMES of a magnitude outside the historical record, must remain a matter of extrapolation from the known events, or a subject for theoretical guidance.
The idea that we are at the mercy  of a scale-free event distribution that allows for the unpredictable occurrence of a ``Black Swan'' superflare had originally been suggested long before the Kepler observations began \cite{1977Natur.268..510W}.
In the meanwhile, evidence has accumulated establishing the existence of a downward break in the distribution of SEP fluences, suggesting a safer extrapolation based mainly on the fossil record -- but we really do not understand the relationship between flare energy and SEPs in this domain.

Theoretical work on flare occurrence remains far from a capability for prediction, even though (not discussed in this review) short-term anticipation of flares does achieve some measure of success on a few-day time scale (e.g., \cite{2012ApJ...747L..41B}).
On longer time scales the occurrence patterns have strong correlations in space and time, such as the solar cycle, but the eruption of magnetic flux otherwise remains a mystery.
The organization of surface flux into spots may not play much of a role in flare occurrence, though of course spots and flares both involve the physics of magnetized plasmas.

This paper has discussed two important new developments related to the occurrence of solar extreme events, namely the successful identifications of transient events in the radioisotope data, on a thousand-year timescale, and the rapidly growing database of ``superflares'' on solar-type stars from \textit{Kepler}.
The radioisotope events clearly have an extraterrestrial origin in some high-energy phenomenon, such as a solar flare.
If in fact solar in origin, these events immediately extend require an extension of the observed distribution to higher event energies, and to suggest less steepening above the observed roll-over fluence.
Similarly the superflare occurrence on solar-type stars also suggests that the roll-over may not be so severe as though.
Neither of these new inputs, though, demands a major change in the basic picture of solar event occurrence, namely that the distribution is a flat one that requires a roll-over or downward break of some sort. 
The nature of the roll-over in the distribution has proven difficult to assess partly because of small-number statistics, but also because of the necessity of linking disparate and ill-understood  datasets in a credible manner.
We have enough information now to make a firmer (and lower) estimate of the probability of a solar super-event, but no formal literature on such an estimate yet seems to exist in the literature.

\ack
I thank NASA for support under contract NAS 5-98033 for RHESSI, as well as the International Space Science Institute (Bern) and the University of Glasgow for hospitality during the preparation of this work.

\section*{References}
\bibliographystyle{iopart-num}
\bibliography{kiel}

\end{document}